\documentclass[aip,graphicx,reprint]{revtex4-1}
\usepackage[version=4]{mhchem}
\usepackage{graphicx}
\usepackage{xcolor}
\draft 

\begin{document}


\title{Direct three-body atom recombination: halogen atoms} 



\author{Rian Koots}
\affiliation{Department of Physics and Astronomy, Stony Brook University}

\author{Grace Ding}
\affiliation{Department of Physics and Astronomy, Stony Brook University}

\author{Jes\'us P\'erez-R\'ios}
\email{jesus.perezrios@stonybrook.edu}
\affiliation{Department of Physics and Astronomy, Stony Brook University}


\date{\today}

\begin{abstract}

The recombination of halogen atoms has been a research topic in chemical physics for over a century. All theoretical descriptions of atom recombination depend on a two-step assumption, where two colliding atoms first form an unstable complex before a third colliding body either relaxes or reacts with it to yield a diatomic molecule. These mechanisms have served well in describing some of the dynamics of atom recombination, but have not yet provided a full theoretical understanding. In this work, we consider the role of the direct three-body recombination mechanism in halogen recombination reactions X + X + M $\rightarrow$ X$_2$ + M, where X is a halogen atom, and M is a rare gas atom. Our results agree well with experimental bromide and iodine recombination measurements, demonstrating that direct three-body recombination is essential in halogen recombination reactions.

\end{abstract} 

\maketitle 

\section{Introduction}

Atom recombination refers to the production of a diatomic molecule and a third free body carrying the excess energy as the result of a termolecular reaction between three atoms. At first glance, the treatment of three reactants may seem intractable, compare with two for bimolecular or one in unimolecular reactions. Despite this, atom recombination became a main subject of interest as modern chemical physics emerged. The first mechanism that was proposed for atom recombination, or generally any termolecular reaction, is the Lindenmann-Hinshelwood mechanism~\cite{Lindemann,Hinshelwood}, introduced in the 1920's. Within this mechanism, atomic recombination occurs via two interleaved two-body collisions. The first collision forms an unstable complex between two colliding atoms, which is then stabilized by a second collision of a third atom. At nearly the same time, Rabinowitch et al. pioneered experimental studies of the atomic recombination reaction X + X + M $\rightarrow$ X$_2$ + M, where X is bromine~\cite{Br1,Br2} and iodine~\cite{I1} in different background gases of atom M. These experiments fueled the community and the interest in atom recombination grew considerably, especially the study of halogen recombination, yielding a considerable literature of experimentally determined termolecular rates~\cite{Blake_1971,Harada_2006,Blake_1970,Strong_1957,Britton_1956,Royal_1953,ip_recombination_1969,christie_i_1953,christie_ii_1955,porter_iii_1961,engleman_iodine_1960,hippler_recombination_1972,hippler_role_1974}.

From a theoretical standpoint, halogen recombination reactions, X + X + M $\rightarrow$ X$_2$ + M, have been assumed to occur via the Lindemann–Hinshelwood mechanism. Depending of the nature of the two-body complex, two different reaction mechanisms have been proposed: the energy transfer (ET) and the radical molecule complex (RMC) mechanisms. The ET mechanism applies when the first bimolecular reaction is between two halogen atoms yielding a complex that is stabilized via the third body M:

\begin{eqnarray}
\ce{X + X  & -> &X^*2 }\\
\ce{X^*2 +  M &-> &X_2 + M}
\end{eqnarray}
On the contrary, the RMC mechanism describes a first collision between a halogen atom X and the third body M, yielding a complex that further reacts with the second halogen atom as 

\begin{eqnarray}
\ce{X + M  & -> &XM^* }\\
\ce{XM^* +  X &-> &X_2 + M}
\end{eqnarray}
These two reaction mechanisms have been extensively studied in the literature~\cite{clarke_trajectory_1971,clarke_trajectory_1972,clarke_trajectory_1973,wong_i2_1973,wong_br2_1973,wong_trajectory_1974,chang_trajectory_1976,burns_trajectory_1980,bunker_mechanics_1960,gelb_nonequilibrium_1972,kim_temperature_1967,rice_recombination_1941,bunker_interpretation_1958,hackmann_recombination_1974} for halogen recombination. However, a general theoretical framework which produces a satisfactory understanding of halogen recombination is still lacking. 


Another route toward halogen recombination is the direct three-body recombination (DTBR), consisting of an almost simultaneous collision of three atoms to form a diatomic molecule. Although this reaction mechanism remains largely unexplored, it has been shown that DTBR provides a satisfactory explanation of several atom recombination problems, including sulfur recombination~\cite{Koots_2024}, helium recombination~\cite{JPR_2014}, ion-atom-atom recombination~\cite{Krukow_2016,Thielemann_2025,JPR_2018,JPR_2024}, molecular anion-atom-atom recombination~\cite{Lochmann_2023,JPR_2024}, dipole-dipole-dipole recombination in the ultracold regime~\cite{Stevenson_2024} and even more complex reactions such as ozone formation~\cite{Mirahmadi_2022}. Therefore, it seems plausible that the DTBR mechanism may play a significant role in halogen recombination reactions.

In this work, we present a DTBR approach for halogen recombination based on a classical trajectory calculation using hyperspherical coordinates. We focus on bromine and iodine recombination in rare gas background gas. The underlying potential energy surface is assumed to result from the three pair-wise interactions between the atoms. Each of the pair-wise interactions is calculated ab initio and ulteriorly fit to a Lennard-Jones potential. Our results show excellent agreement independently of the rare gas atom and the temperature range. Therefore, direct three-body recombination should not be neglected from halogen atom recombination, and not only that, it should be embraced in order to understand halogen recombination fully.



\section{Potential Energy Curves}
The interaction potential for X$_2$M molecules, where X=I and Br and \ce{M} = [\ce{He}, \ce{Ne}, \ce{Ar}, \ce{Kr}, \ce{Xe}], is assumed to be described by pairwise interactions. The potentials related to the X$_2$ molecules, $V_{X_2}(r)$, are well known and are taken from reference data~\cite{c6coeffs2016,Li_Thompson_2004,Liu_2024}. Those related to the XM molecule, $V_{XM}(r)$, are calculated via high level electronic structure methods. Specifically, we calculate ab initio potential energy curves for each pair of interactions \ce{IM} and \ce{BrM}, calculated at the coupled-clusters with singles, doubles and perturbative triples excitations CCSD(T), as implemented in \texttt{MOLPRO} quantum chemistry program~\cite{MOLPRO_brief}. The aug-cc-pVQZ basis set was used for noble gas \ce{M}~\cite{basis_stes} and the def2-QZVPPD basis set for \ce{I}, \ce{Br}~\cite{def2_basis}. Both of basis set considered include polarization effects able to accurately describe long range interactions between atoms within the molecule.

\begin{figure}[h!]
    \includegraphics[width=0.5\textwidth]{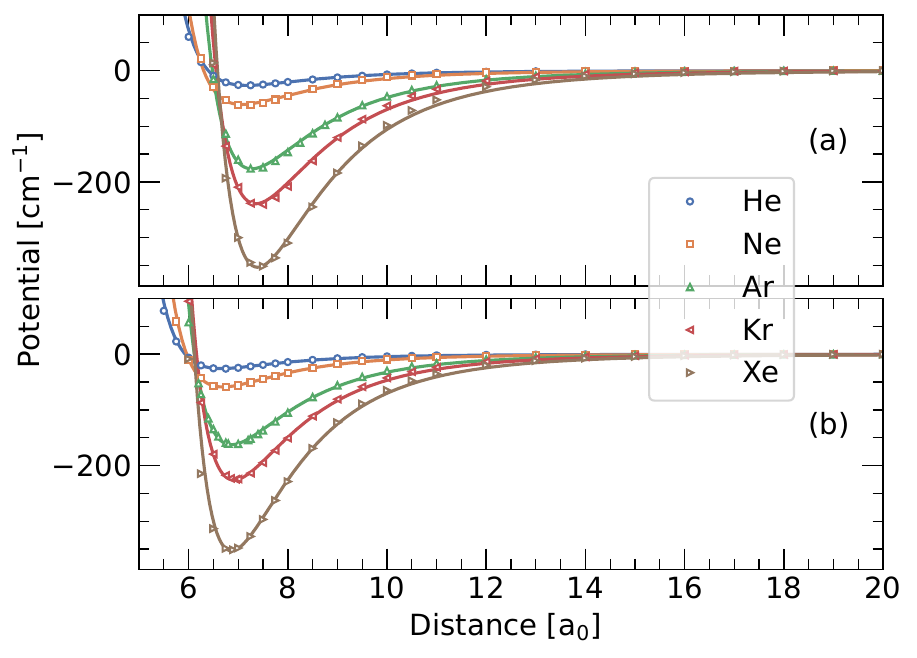}
    \caption{\label{fig:potentials} Potential energy curves. The symbols represent the CCSD(T) calculations for (a) I-M and (b) Br-M with \ce{M} = [\ce{He}, \ce{Ne}, \ce{Ar}, \ce{Kr}, \ce{Xe}, whereas the solid line is their fitting to a Lennard-Jones potential.}
\end{figure}

Since MX are van der Waals molecules, their ab initio interaction potential can be efficiently fitted  by the Lennard-Jones potential $V(r) = \frac{C_{12}}{r^{12}} - \frac{C_6}{r^6}$. On the contrary, X$_2$ are covalent molecules which are typically described via a Morse potential. However, since the essential properties relevant for three-body recombination are the long-range interaction, given by $C_6$, and the dissociation energy $D_e$~\cite{Mirahmadi_2021}, we fit the halogen molecules with Lennard-Jones potential as well. We use $C_6$ and $D_e$  as fitting parameters to each potential, setting $C_{12} = C_{6}^2/(4D_e)$. The parameters to the fitted potentials are shown in Table~\ref{tab:lj_params}. A comparison between the Lennard-Jones potentials and the ab initio potential energy curves are displayed in Fig.~\ref{fig:potentials}, where it is noticed that the Lennard-Jones potential indeed describes the interaction energies of MX molecules well. 




\begin{table}[h]
    \centering
    \setlength{\tabcolsep}{10pt} 
    \begin{tabular}{cccc}
        \hline
        Species & $C_6$ (a.u.) & $C_{12}$(10$^{6}$ a.u.) & D$_e$ (cm$^{-1}$) \\
        \hline
        \hline
        I-I   & 389$^{(a)}$    & 0.67  & 12403$^{(b)}$   \\
        I-He  & 33.2 (0.2)  & 2.3 (0.1) & 26.47 (0.07)   \\
        I-Ne  & 70 (2)  & 4 (1) & 61 (1)   \\
        I-Ar  & 240 (2) & 18 (1) & 178 (1)   \\
        I-Kr  & 337 (4) & 26 (2) & 240 (2)   \\
        I-Xe  & 524 (8) & 43 (6) & 353 (4)   \\ 
        
        Br-Br & 187$^{(a)}$    & 0.12 & 16057$^{(c)}$   \\
        Br-He & 20.4 (0.1)  & 0.89 (0.04) & 25.8 (0.1)   \\
        Br-Ne & 48.1 (0.3)  & 2.2 (0.1) & 58.9 (0.3)   \\
        Br-Ar & 155 (1) & 8.1 (0.4) & 163 (1)  \\
        Br-Kr & 225 (2)   & 12.3 (0.9) & 226 (1)   \\
        Br-Xe & 337 (3)   & 18 (1) & 354 (3)   \\
        \hline
        \hline
    \end{tabular}
    \caption{Parameters used for the Lennard-Jones potential. For the X-M potentials, C$_6$ and D$_e$ were used as fitting parameters to CCSD(T) calculations. The fitting errors are quoted in parenthesis. Note that the errors in $C_{12}$ are due to the propagation from its relationship via $ C_{12} = C_{6}^2/(4D_e)$.  The X-X parameters were obtained from (a) [Ref.~\cite{c6coeffs2016}] (b)[Ref.~\cite{Li_Thompson_2004}] and (c) [Ref.~\cite{Liu_2024}]. }
    \label{tab:lj_params}
\end{table}

\section{Theoretical approach}
We use the Py3BR~\cite{py3br_2024} software package to perform the DTBR calculations, which utilizes classical trajectories in hyperspherical coordinates to calculate reactive rate constants. The details of the DTBR method have been outlined previously~\cite{Mirahmadi_rev2022}, which uses a Monte Carlo approach to calculate the energy-dependent three-body recombination rate constant as defined in the six-dimensional hyperspherical coordinate system:
\begin{equation}\label{eq:rate_e}
    k_3(E_c) = \frac{8\pi^2}{3}\sqrt{\frac{2E_c}{\mu}}\int_0^{b_{\text{max}}}\mathcal{P}(E_c,b)b^4db 
\end{equation}
where $\mu = \sqrt\frac{m_X m_X m_M}{m_X + m_X + m_M}$ is the three-body reduced mass, $E_c$ is the collision energy, $b$ is the impact parameter, and $\mathcal{P}(E_c,b)$ is the probability of a recombination event to occur, the so-called opacity function. For a given collision energy, we use a set of equally spaced impact parameters that span up to a maximum value $b_{\text{max}}$ such that $\mathcal{P}(E_c,b > b_{\text{max}}) = 0$. The recombination probability for a reactive collision at a given collision energy and impact parameter is simply the ratio of reactive trajectories, $n_{\text{R}} (E_c,b)$, to the total number of trajectories run, $n_T(E_c,b)$, with a counting error representing one standard deviation of uncertainty: 
\begin{equation}
    P(E_c,b) = \frac{n_{\text{R}}(E_c,b)}{n_T(E_c,b)} \pm \frac{n_{\text{R}}(E_c,b)}{n_T(E_c,b)}\sqrt{\frac{n_T(E_c,b)-n_{\text{R}}(E_c,b)}{n_{\text{R}}(E_c,b)n_T(E_c,b)}}
\end{equation}

\subsection{Mass Weighted Coordinates}
Our DTBR method relies on a coordinate transformation from three dimensions (3D) to six dimensions (6D) in order to define the initial conditions of a collision and satisfy the conditions for the rate coefficient integral (Eq.~\ref{eq:rate_e}). The methodology has been presented elsewhere~\cite{JPR_2014,perezrios2020} and recently reviewed~\cite{Mirahmadi_rev2022}. The initial 6D position ($\vec{\rho}_{6D}$) and momentum ($\vec{P}_{6D}$) vectors are constructed from the 3D Jacobi position vectors ($\vec{\rho}_1$, $\vec{\rho}_2$) and momentum vectors ($\vec{P}_1$, $\vec{P}_2$), which define the relative distances and associated conjugate momenta. For an X + X + M collision, $(\vec{\rho}_1, \vec{P}_1)$ are position and momentum vectors between X-X, and $(\vec{\rho}_2, \vec{P}_2)$ are position and momentum vectors between M and the center of mass of X-X.

This coordinate transformation is evident in the transition from the 3D Hamiltonian to the 6D Hamiltonian. For an X + X + M collision, in the 3D Jacobi coordinate system, the Hamiltonian is written as
\begin{equation}
    H = \frac{\vec{P}_{1}^2}{2\mu_{XX}} + \frac{\vec{P}_{2}^2}{2\mu_{MXX}} + V(\vec{\rho}_{1},\vec{\rho}_{2}),
\end{equation}
where $\mu_{XX} = (1/m_X + 1/m_X)^{-1}$ and $\mu_{MXX} = \left(1/(m_X + m_X) + 1/m_M\right)^{-1}$. Due to the large mass differences of some X-M collisions, we use a mass-weighted transformation to describe the 6D position and momentum vectors,
\begin{align}
    \vec{\rho}_{6D} &= \left(\sqrt{\frac{\mu_{XX}}{\mu}}\vec{\rho}_{1}, \sqrt{\frac{\mu_{MXX}}{\mu}}\vec{\rho}_{2} \right) \\
    \vec{P}_{6D} &= \left(\vec{P}_{1}, \vec{P}_{2} \right),
\end{align}
where $\mu = \sqrt{m_Mm_X^2/(m_M+2m_X)}$. In the 6D coordinate system, the Hamiltonian is
\begin{equation}
    H = \frac{\vec{P}_{6D}}{2\mu} + V(\vec{\rho}_{6D}).
\end{equation}

\section{Results and Discussion}
\begin{figure}[h!]
    \centering
    \includegraphics[width=\linewidth]{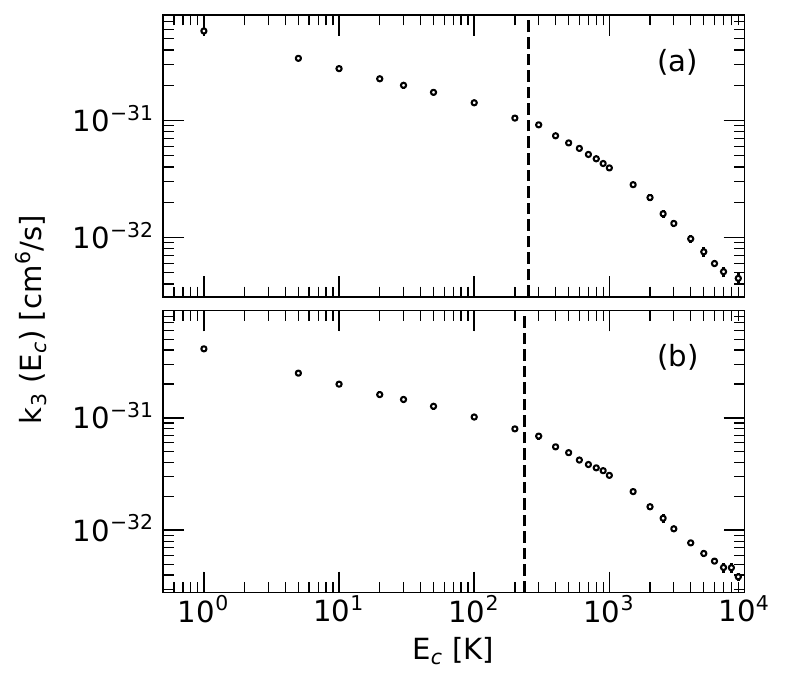}
    \caption{\label{fig:k3_Ar} Energy-dependent recombination rates of (a) \ce{I2} and (b) \ce{Br2} in the presence of \ce{Ar}. The dashed vertical line represents the dissociation energy of \ce{IAr} and \ce{BrAr} obtained via the Lennard-Jones fit to the reported CCSD(T) pairwise interaction.}
\end{figure}

Among all halogen recombination reactions, X + X + Ar $\rightarrow$ X$_2$ + Ar has received special attention, and it has been used as a benchmark for theoretical models~\cite{bunker_mechanics_1960,wong_i2_1973,clarke_trajectory_1971,clarke_trajectory_1972,clarke_trajectory_1973,wong_trajectory_1974,wong_br2_1973,chang_trajectory_1976,burns_trajectory_1980}. Therefore, we start by exploring halogen recombination mediated by Ar. The energy-dependent recombination rate coefficients given by Eq.~(\ref{eq:rate_e}) are shown in Fig.~\ref{fig:k3_Ar}. These results were calculated using 30,000 trajectories per impact parameter over a range of 26 collision energies. As is typical with three-body recombination studies~\cite{Mirahmadi_rev2022}, a change of trend is seen near the dissociation energy of the X-M interaction (vertical dashed-line). Below this dissociation energy, in the ``low-energy'' regime, three-body recombination reactions exhibit a clear power law behavior according to the long-range interaction potential $C_6/r^6$. Deviation from this trend occurs at energies larger than the X-M dissociation energy, showing a steeper decline in the rate coefficient~\cite{Mirahmadi_rev2022}. 

The thermally averaged rate constant is obtained by integrating the energy-dependent rate $k_3$ ($E_c$), over the appropriate three-body Maxwell-Boltzmann distribution of collision energies:
\begin{equation}
    k_3(T) = \frac{1}{2(k_BT)^3}\int_0^{\infty}k_3(E_c)E_c^2e^{-E_c/(k_BT)}dE_c.
\end{equation}
where $k_B$ is the Boltzmann constant and $T$ is the temperature of the system.

\subsection{Post-collision survival and degeneracy factors}
 
 In general, molecules resulting from atom recombination appear in weakly bound vibrational states ~\cite{Mirahmadi_2022}. Subsequent collisions with surrounding atoms will either remove energy from the molecule resulting in a more deeply bound state, or dissociate the molecule resulting in a loss. This post-collision loss can be represented as a statistical survival factor $\Delta E/k_BT$~\cite{Troe}, where $T$ is the temperature and $\Delta E$ is the average energy spacing of the \ce{X2} molecule. This factor applies for when $T > \Delta E$, and at lower temperatures the survivability factor is 1. The average energy spacing is calculated by considering the density of vibrational states for each molecule according to their long-range interaction. For \ce{I2}, we use $\Delta E$ = 118.2 cm$^{-1}$ (179 K) and for \ce{Br2} we use $\Delta E$ = 209.9 cm$^{-1}$ (302 K). This approach has been used previously, for example, in the study of ozone formation, yielding accurate results~\cite{Mirahmadi_2022}. 

In addition to the post-collision survivability factor, we include an electronic degeneracy factor $g$, due to the splitting of the ground state of each of the $^2$P$_{\frac{3}{2}}$ X atoms into 4 components~\cite{bunker_mechanics_1960}. Of the 16 possible states, we consider only the attractive $^1\Sigma$ and $^3\Pi$ states as responsible for recombination reactions, yielding $g = 5/16$.  

After considering both the post-collision quenching and degeneracy of electronic states in the recombination of \ce{I2} and \ce{Br2}, we report the final recombination rate coefficient as:
\begin{equation}
    k_{\text{rec}}(T) = gk_3(T)\frac{\Delta E}{k_BT}
\end{equation}

\begin{figure}[h!]
    \centering
    \includegraphics[width=\linewidth]{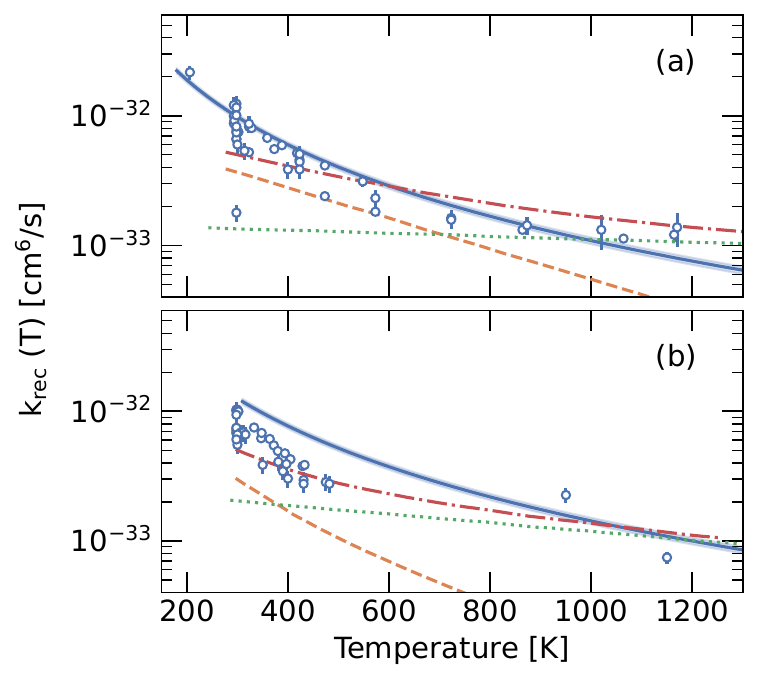}
    \caption{ The recombination rate coefficient of \ce{I2} (top) and \ce{Br2} (bottom) in a \ce{Ar} bath gas as a function of temperature. For both plots, the solid blue lines are rates predicted from the DTBR method, with the associated errors in shaded blue. The green dotted lines show the predicted rate corresponding to the ET mechanism, orange dashed lines show the predicted rate from the RMC mechanism, and the red dash-dotted lines show the sum of the two mechanisms~\cite{wong_i2_1973,wong_br2_1973}. The blue open circles represent experimental data~\cite{baulch_1981}.}
    \label{fig:kt_Ar} 
\end{figure}

The DTBR calculated $k_{\text{rec}}(T)$ for third body M$=$\ce{Ar} is shown in Fig.~\ref{fig:kt_Ar} for both \ce{I2} and \ce{Br2} recombination. We see excellent agreement with experimental data, capturing the correct order of magnitude and temperature dependence of the recombination rate constant, suggesting that we have captured the correct physics. For comparison, we also show the rates calculated by classical trajectory studies of the ET and RMC mechanisms for which M=\ce{Ar}~\cite{wong_i2_1973,wong_br2_1973}, which demonstrate the roles of each mechanism in atom recombination. Our results indicate that DTBR is indeed a relevant reaction mechanism for the recombination of iodine and bromine atoms in argon.  

\begin{figure}[h!] 
    \centering
    \includegraphics[width=\linewidth]{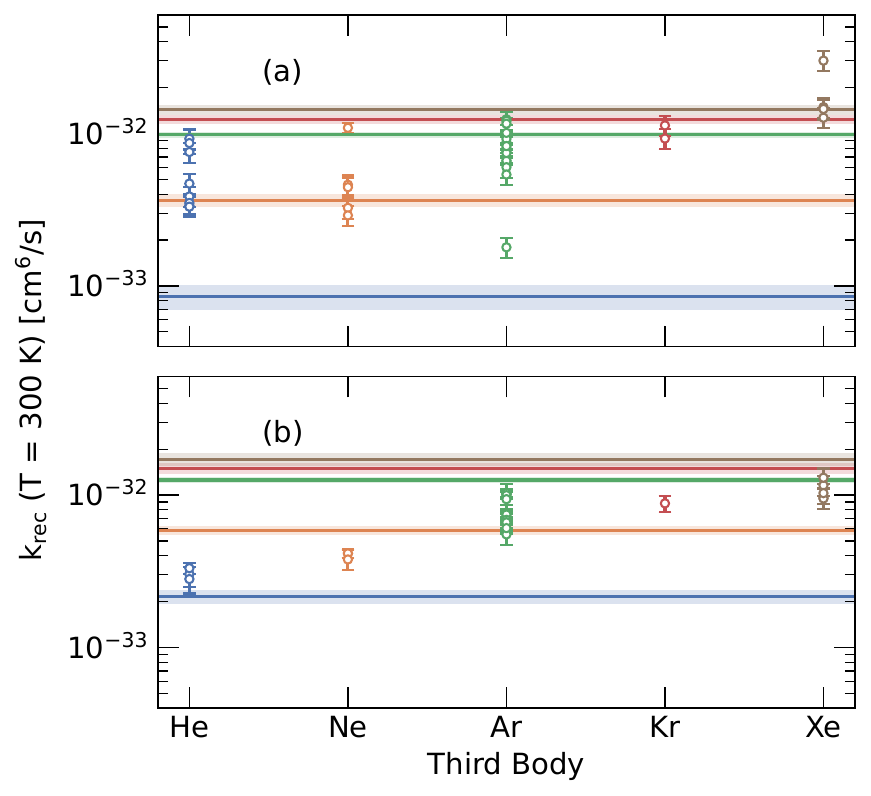}
    \caption{Atomic recombination rates for (a) \ce{I2} and (b) \ce{Br2} in bath gases of M = He (blue), Ne (orange), Ar (green), Kr (red), and Xe (brown). Experimental rates~\cite{baulch_1981} were determined at temperatures ranging from 280 K to 320 K, and are shown as open circles with errorbars. The DTBR calculated rates at T = 300~K are shown by horizontal lines, with statistical errors given by the shaded region.}
    \label{fig:kt_all}
\end{figure}

We perform the same procedure for the recombination of I and Br in bath gases of He, Ne, Kr, and Xe. Since most of the experimental data is available at or near room temperature across all combinations of species, we show the thermally averaged rate coefficients at $T$ = 300~K alongside experimentally determined rates in the temperature range of 280~K to 320~K in Fig.~\ref{fig:kt_all}. Here, we see that the DTBR mechanism properly predicts the rate constant as a function of the nature of the third body. The agreement is remarkable, independent of the halogen atom. These results imply that the underlying interaction potential is accurate enough to describe the three-body dynamics and that DTBR is an efficient reaction mechanism for atom recombination at room temperature. However, we notice some deviations when the third body is He.

\begin{figure}[h!]
    \centering
    \includegraphics[width=\linewidth]{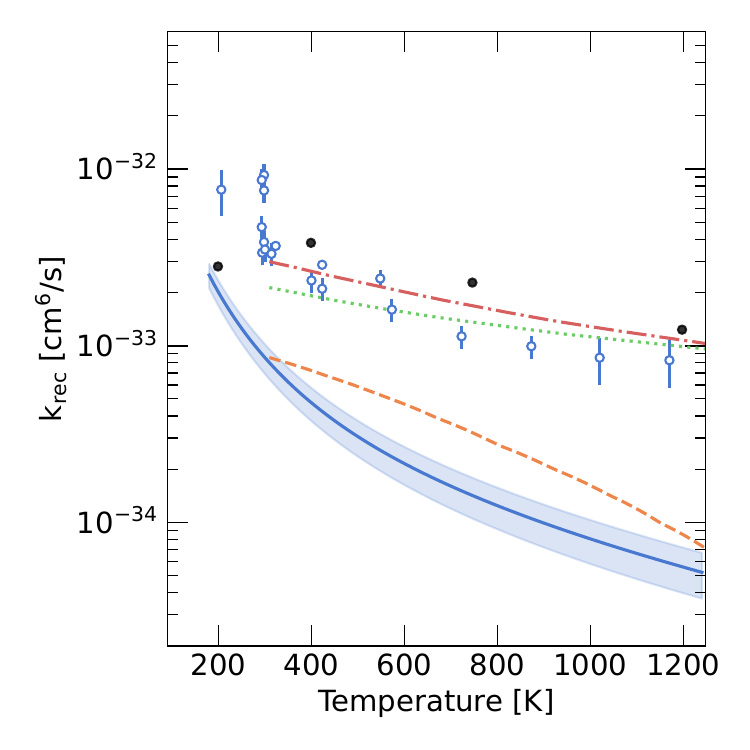}
    \caption{The recombination rate coefficient of \ce{I2} in a \ce{He} bath gas as a function of temperature. The solid line represents the thermally averaged recombination rate coefficient calculated from DTBR.  The green dotted and orange dashed lines show the rates calculated by a classical trajectory study of the ET and RMC mechanisms, respectively, while the red dash-dotted line is the sum of the two.~\cite{wong_i2_1973}. The black closed circles show  
    further theoretical investigations into the validity of the ET mechanism in atomic recombination reactions~\cite{burns_trajectory_1980}. The experimentally measured rates~\cite{baulch_1981} are shown in blue open circles.}
    \label{fig:kt_he}
\end{figure}

To investigate further, we plot the temperature-dependent recombination rate coefficient of the I + I + He system over the experimental range in Fig.~\ref{fig:kt_he}. For comparison, we include results from classical trajectory studies of the ET and RMC mechanisms~\cite{wong_i2_1973} which were found to be consistent with experimental data. We also include results from a subsequent study which investigated the role of the predominant ET mechanism in the \ce{I + I + He} reaction~\cite{burns_trajectory_1980}.The improved study shows a negative slope of the atom recombination rate as a function of the temperature between 200~K and 400~K, opposing the experimental results. To this point, the authors conclude that "Whenever the ET mechanism predominates, it would seem more appropriate to substitute it with a more rigorous three body problem solution." On the contrary, we see that, although the DTBR predicted rate differs in magnitude, it properly describes the temperature-dependent behavior of \ce{I2} recombination, suggesting that the DTBR mechanism plays a significant role in the recombination of \ce{I2} in \ce{He}. One possible source of error in the magnitude determination is the survivability factor, which may significantly vary across systems.
More specific determination of these factors may require further investigation into the quantum mechanical properties of these collisions. 





\section{Conclusion}
To study the DTBR mechanism, we employ classical trajectories using hyperspherical coordinates to calculate the recombination rate constants of \ce{I} and \ce{Br} in various rare gases, including \ce{He}, \ce{Ne}, \ce{Ar}, \ce{Kr}, and \ce{Xe}. We present new ab initio potential curves between each species calculated at the CCSD(T) level of theory using a large basis set with polarization effects to efficiently describe the long-range tail of the interaction potential, essential for recombination reactions.  After fitting these curves to a Lennard-Jones potential, energy-dependent recombination rate coefficients are calculated. These energy-dependent rates are then thermally averaged, where post-collision dissociation and electronic degeneracy factors were considered. With these factors in place, we find excellent agreement in the temperature-dependent atom recombination rates across all systems. 

Investigation into the nature of atom recombination was fueled by numerous experiments and supporting theories throughout the 20th century focused on halogen recombination in a rare gas bath. The framework for treating such reactions has been to employ a two-step mechanism such as the energy transfer mechanism or the radical molecule complex mechanism. Analysis of atom recombination in this way requires careful consideration of the system at hand, since the importance of each mechanism and their parameters varies across systems. While this approach has yielded accurate results, it has not provided a general theoretical description of the recombination of atoms. Instead, we show that the direct three-body recombination (DTBR) mechanism, in which the reaction occurs without invoking an intermediate complex, is an alternative reaction mechanism to study the recombination of halogens in rare gas atoms. Using this approach, we find excellent agreement with experiments across any bath species, providing a first principle description for atom recombination. 

\section{Acknowledgments}

The authors acknowledge the generous support of the Simons Foundation.



%
%

%


\section*{Author Declarations}
\subsection*{Conflict of Interest}
The authors have no conflicts to disclose.

\subsection*{Author Contributions}

\section*{Data Availability Statement}
The data that supports the findings of this study are available within the article [and its supplementary material].

\bibliography{references}

\end{document}


\widetext
\begin{center}
\textbf{\large Supplemental Materials: Direct three-body atom recombination: halogen atoms}
\end{center}

\makeatletter
\renewcommand{\theequation}{S\arabic{equation}}
\renewcommand{\thefigure}{S\arabic{figure}}
\renewcommand{\thetable}{S\arabic{table}}
\renewcommand{\bibnumfmt}[1]{[S#1]}
\renewcommand{\citenumfont}[1]{S#1}

\section{Potential Energy Curves}
The pairwise potentials calculated at the CCSD(T) level for each I-M and Br-M pair studied in this work are shown in Table~\ref{tab:pairwise_I} and Table~\ref{tab:pairwise_Br}.  
\begin{table}[h]
    \centering
    \setlength{\tabcolsep}{10pt} 
    \begin{tabular}{ccccccccccc}
\hline
\hline

 $r$ [a$_0$] & V$_{\ce{IHe}} (r)$ & V$_{\ce{INe}} (r)$ & V$_{\ce{IAr}} (r)$ & V$_{\ce{IKr}}(r)$ & V$_{\ce{IXe}} (r)$ \\
\hline
4.25 & 3452 & 7807 & 20920 & 29630 & 44220 \\
4.50 & 2182 & 4855 & 13300 & 18840 & 28060    \\
4.75 & 1357 & 2972 & 8349 & 11840 & 17540     \\
5.00 & 825 & 1779 & 5155 & 7319 & 10760  \\
5.25 & 485.50 & 1029 & 3108 & 4428 & 6441 \\
5.50 & 271.70 & 564.80 & 1804 & 2591 & 3726 \\
5.75 & 139.80 & 282.80 & 983.2 & 1433 & 2039 \\
6.00 & 60.73 & 116.2 & 473.5 & 712.2 & 1006  \\
6.25 & 15.32 & 21.68 & 164.7 & 271.3 & 381.6 \\
6.50 & -9.174 & -28.82 & -15.17 & 10.53 & 13.96 \\
6.75 & -21.07 & -52.94 & -113.2 & -135.6 & -193.2 \\
7.00 & -25.7 & -61.78 & -160.0 & -209.4 & -300.0 \\
7.25 & -26.34 & -62.11 & -175.9 & -238.3 & -344.7 \\
7.50 & -25.00 & -58.05 & -173.9 & -240.8 & -351.5 \\
7.75 & -22.83 & -52.06 & -162.4 & -228.3 & -336.5 \\
8.00 & -20.43 & -45.52 & -146.5 & -208.3 & -309.9 \\
8.5 & -15.93 & -33.36 & -112.7 & -162.2 & -245.0  \\
9.00 & -12.22 & -23.81 & -84.06 & -120.3 & -184.0 \\
9.50 & -9.24 & -16.86 & -62.31 & -87.35 & -134.9  \\
10.0 & -6.913 & -11.92 & -46.48 & -62.99 & -98.13  \\
10.5 & -5.136 & -8.494 & -34.94 & -45.65 & -71.46  \\
11.0 & -3.819 & -6.123 & -26.45 & -33.36 & -52.37  \\
12.0 & -2.129 & -3.314 & -15.34 & -18.44 & -28.93  \\
13.0 & -1.229 & -1.887 & -9.02 & -10.53 & -16.59     \\
14.0 & -0.7023 & -1.119 & -5.333 & -6.145 & -9.745   \\
15.0 & -0.417 & -0.6804 & -3.160 & -3.731 & -5.772  \\
16.0 & -0.2414 & -0.3951 & -1.844 & -2.195 & -3.402  \\
17.0 & -0.1317 & -0.2414 & -1.032 & -1.097 & -1.931  \\
18.0 & -0.06584 & -0.1097 & -0.5267 & -0.6584 & -1.01 \\
19.0 & -0.02195 & -0.04389 & -0.2195 & -0.2195 & -0.3951 \\
20.0 & 0 & 0 & 0 & 0 & 0  \\
\hline
\hline
\end{tabular}
\caption{I-M pairwise potentials calculated at the CCSD(T) level using the aug-cc-PVQZ basis set for M = He, Ne, Ar, Kr, Xe and the def2-QZVPPD basis set for I. The first column is the interatomic distance $r$ in a$_0$. The remaining columns, V$_{\ce{IM}} (r)$, are CCSD(T) calculated energies for each \ce{IM} species pair at distance $r$, reported in cm$^{-1}$.}
\label{tab:pairwise_I}

\end{table}
\clearpage
\begin{table*}[h]
    \centering
    \setlength{\tabcolsep}{10pt} 
    \begin{tabular}{ccccccccccc}
\hline
\hline

 $r$ [a$_0$] & V$_{\ce{BrHe}} (r)$ & V$_{\ce{BrNe}}(r)$ & V$_{\ce{BrAr}}(r)$  & V$_{\ce{BrKr}}(r)$ & V$_{\ce{BrXe}}(r)$ \\
\hline
4.25 & 1850 & 4536 & 11970 & 16540 & 23540 \\
4.50 & 1097 & 2667 & 7305 & 10040 & 13970 \\
4.75 &  630.80 & 1523 & 4371 & 5982 & 8043 \\
5.00 &  346.30 & 833.10 & 2538 & 3470 & 4448 \\
5.25 &  176.70 & 424.70 & 1402 & 1927 & 2318 \\
5.50 &  78.13  & 189.40 & 709.10 & 987.90 & 1083 \\
5.75 &  23.04 & 58.82 & 295.6 & 424 & 381 \\
6.00 &  -6.145 & -10.1 & 57.5 & 95.25 & -8.34 \\
6.25 &  -19.97 & -43.24 & -71.55 & -87.13 & -214.6 \\
6.50 & -25.24 & -56.62 & -134.5 & -179.5 & -313.4 \\
6.75 &  -25.9 & -59.04 & -158.9 & -217.7 & -349 \\
7.00 &  -24.36 & -56.19 & -161.1 & -224.3 & -348.3 \\
7.25 &  -21.73 & -50.7 & -151.9 & -213.8 & -327.5 \\
7.50 &  -19.09 & -44.55 & -137.2 & -194.9 & -296.9 \\
7.75 &  -16.46 & -38.63 & -120.9 & -172.9 & -262.9 \\
8.00 &  -14.05 & -33.14 & -104.9 & -150.8 & -228.9 \\
8.5  &  -10.1 & -24.14 & -77.04 & -111.3 & -169.2 \\
9.00 &  -7.243 & -17.34 & -55.97 & -80.77 & -122.9 \\
9.50 &  -5.267 & -12.51 & -40.6 & -58.6 & -89.11 \\
10.0 &  -3.731 & -9.218 & -29.63 & -42.58 & -64.75 \\
10.5 &  -2.853 & -6.584 & -21.95 & -31.38 & -47.41 \\
11.0 &  -1.975 & -4.828 & -16.24 & -23.26 & -35.12 \\
12.0 &  -1.097 & -2.853 & -9.218 & -13.17 & -19.75 \\
13.0 &  -0.6584 & -1.536 & -5.487 & -7.682 & -11.63 \\
14.0 &  -0.2195 & -0.8779 & -3.292 & -4.609 & -7.023 \\
15.0 &  -0.2195 & -0.4389 & -1.975 & -2.634 & -4.17 \\
16.0 &  0 & -0.2195 & -1.097 & -1.536 & -2.414 \\
17.0 &  0 & -0.2195 & -0.6584 & -0.8779 & -1.317 \\
18.0 & 0 & 0 & -0.4389 & -0.4389 & -0.6584 \\
19.0 & 0 & 0 & -0.2195 & -0.2195 & -0.2195 \\
20.0 &  0 & 0 & 0 & 0 & 0 \\
\hline
\hline
\end{tabular}
\caption{Br-M pairwise potentials calculated at the CCSD(T) level using the aug-cc-PVQZ basis set for M = He, Ne, Ar, Kr, Xe and the def2-QZVPPD basis set for Br. The first column is the interatomic distance $r$ in a$_0$. The remaining columns, V$_{\ce{BrM}} (r)$, are CCSD(T) calculated energies for each \ce{BrM} species pair at distance $r$, reported in cm$^{-1}$.}
\label{tab:pairwise_Br}

\end{table*}



\begin{figure}[h]
    \centering
    \includegraphics[width=0.5\linewidth]{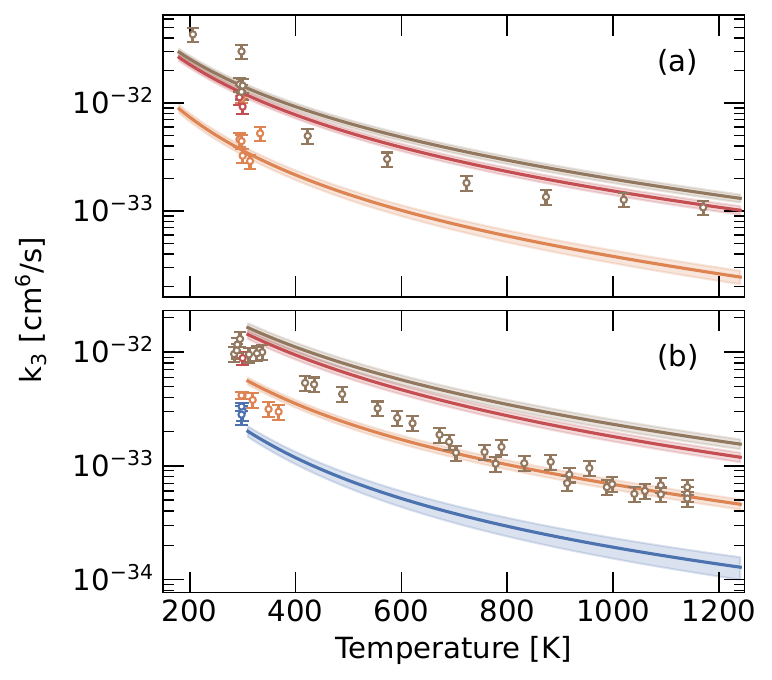}
    \caption{Thermally averaged recombination rates for (a) \ce{INe} (orange), \ce{IKr} (red), and \ce{IXe} (brown) and for (b) \ce{BrHe} (blue), \ce{BrNe} (orange), \ce{BrKr} (red), and \ce{BrXe}. Experimental values for each species were obtained from [Ref.~\cite{baulch_1981}], except for \ce{BrXe} which are from [Ref.~\cite{clarke_trajectory_1971}]}
    \label{fig:kt_NeKrXe}
\end{figure}

\section{More temperature-dependent rates}
The thermally averaged three-body recombination rate has been calculated over a large range of temperatures for each X-M pair, of which only \ce{IAr}, \ce{BrAr}, and \ce{IHe} are shown in the main text. In Fig.~\ref{fig:kt_NeKrXe}, we present DTBR calculated rates for the rest of the reactions mentioned, namely \ce{INe}, \ce{IKr}, \ce{IXe}, \ce{BrHe}, \ce{BrNe}, \ce{BrKr}, and \ce{BrXe}. Here, we see that the temperature dependent trend for the recombination rate constant is well captured by the DTBR method with our calculated pairwise potentials across all systems.

\newpage
\section{References}
\bibliography{refs}